\documentclass[useAMS,usenatbib]{mn2e}
\usepackage{natbib}
\usepackage{graphicx}

\title[Evolution of faint radio sources]{Evolution of faint
  radio sources in the VIDEO-XMM3 field}
\author[K. McAlpine, M.J.Jarvis and D.G. Bonfield]{K.McAlpine$^{1}$, M.J.Jarvis$^{1,2}$ and D.G.Bonfield$^{3}$\\
$^{1}$Department of Physics, University of the Western Cape, Private
Bag X17,
Bellville, 7537, South Africa\\
$^{2}$Department of Physics, University of Oxford, Denys Wilkinson Building, Keble Road, OX1 3RH, UK\\
$^{3}$School of Physics, Astronomy and Mathematics, University of
Hertfordshire, College Lane, Hatfield AL10 9AB, UK}
\begin{document}

\date{}

\pagerange{\pageref{firstpage}--\pageref{lastpage}} \pubyear{2013}

\maketitle

\label{firstpage}

\begin{abstract}
It has been speculated that low luminosity radio-loud AGN have the
potential to serve as an important source of AGN feedback, and may be responsible for suppressing star-formation activity
in massive elliptical galaxies at late times. As such the
cosmic evolution of these sources is vitally important to
understanding the significance of such AGN feedback processes and
their influence on the global star-formation history of the universe.



In this paper we  present a new investigation of the evolution of faint radio sources
out to $z{\sim}2.5$. We combine a 1~square degree VLA radio survey,
complete to  a depth of 100~$\mu$Jy, with accurate 10 band photometric
redshifts from the VIDEO and CFHTLS surveys.  The results indicate that
the radio population experiences mild positive evolution out to
$z{\sim}1.2$ increasing their space density by a factor of $\sim$3,
consistent with  results of several previous studies. Beyond $z$=1.2 there is evidence of a slowing down of this
evolution. Star-forming galaxies drive the more rapid evolution at low
redshifts, $z{<}$1.2, while more slowly evolving AGN populations dominate at higher
redshifts resulting in a decline in the evolution of the radio
luminosity function
at $z{>}$1.2. The evolution is best fit by pure luminosity evolution
with star-forming galaxies evolving as $(1+z)^{2.47\pm0.12}$ and AGN as $(1+z)^{1.18\pm0.21}$.  
\end{abstract}


\section{Introduction}

The study of the evolution of faint radio sources has taken on new
significance in recent years due to the realisation that AGN 
play an important role in shaping the star formation
properties of the galaxies they inhabit \citep[for a review see ][]{cattaneo}. As it is now believed that every
galaxy harbours a central supermassive black hole this interplay between
AGN accretion and star-formation, as well as the cosmic evolution of AGN, have become vitally important to the process of understanding galaxy
formation and evolution. Evidence which points to a link between these processes
 include  the tight correlations found between black hole mass and
 both stellar bulge mass \citep{Korm1995,Magorrian1998,MD2002} and
 velocity dispersion \citep{Geb2000,Tremaine2002,Ferr} in galaxies in
 the local universe.  While on cosmological scales the increase in
 star-formation density at higher redshifts is accompanied by a
 corresponding increase in the number density of actively accreting
 supermassive black holes \citep{MRD2007}.  Furthermore 
 incorporating AGN `feedback'  in semi-analytic models of galaxy
 formation produces galaxies whose predicted properties  are in
 better agreement with observations
 \citep{croton2006,Bower2006} particularly at the bright end of the
 luminosity function.  Despite these promising indications 
 many open questions remain about the nature and significance of such postulated AGN
 `feedback' processes and how they proceed as a function of cosmic
 time. 
 
At $\mu$Jy levels radio surveys become increasing dominated
  by star-forming galaxies and  low luminosity AGN \citep[e.g.][]{Simpson2006,Seymour2007,Huyhn2008,Smol2008,Padovani2009}.
As radio waves are unaffected by dust extinction future deep radio
surveys can overcome many of the selection biases present in optical
and X-ray surveys and offer new insights into the co-evolution of these
populations out to high redshifts.  

Efforts to clarify the nature of the $\mu$Jy radio population
have already
revealed that its composition is more
complex than previously believed. It has become widely accepted that 
radio-loud AGN  are in fact powered by two
fundamentally different modes of accretion
(\citealp[e.g.][]{Evans2006}; \citealp*{Hardcastle2007}). The most powerful radio-loud AGN are typically powered by radiatively efficient accretion of
cold gas onto a
geometrically thin, optically thick accretion disk \citep[e.g.][]{Shakura1973,Novikov}. This type of
accretion powers optical and X-ray selected quasars, and produces high
excitation emission lines in the host galaxy spectra. Consequently
radio sources powered by this mode of accretion are referred to as
High Excitation Radio Galaxies (HERGs). Low-luminosity AGN are more frequently powered
by radiatively inefficient \citep{Best2012} accretion of warm gas \citep{Hardcastle2007} from the
intergalactic medium onto a geometrically
thick accretion disk. These  Low
Excitation Radio Galaxies (LERGs) do not exhibit accretion related
X-ray emission \citep*{Hardcastle2006} or mid-infrared emission from an
obscuring nuclear torus \citep*{Ogle2006}, which is
postulated to hide this X-ray emission from view at certain
orientations in quasars \citep{Urry}.  Following the terminology of
\citet{croton2006} these two modes of accretion are 
referred to as `quasar' and `radio' mode respectively. While LERGs and
HERGs are present at all luminosities,  LERGS are more prevalent at
low radio luminosities with HERGS becoming the dominant population at
luminosities  $>$10$^{26}$~W~Hz$^{-1}$ (\citealp*{Hardcastle2009}; \citealp{Herbert2010,Herbert2011,Best2012,Hardcastle2013}),  

Further complications arise from the discovery that in addition to star-forming galaxies and radio-loud AGN, the $\mu$Jy
radio population also contains a significant contribution from sources
that traditionally were classified as radio-quiet
AGN
\citep{Jarvis2004,Simpson2006,Wilman2008,Padovani2009,Wilman2010},
  where the radio emission is much fainter than typical for radio loud
  AGN, i.e. in the case of quasars. They exhibit a radio to
optical light ratio $<$10 \citep[see e.g.][]{Kellermann1989} and are powered by a radiatively efficient accretion mode
similar to that in HERGs.  Currently the contribution from radio-quiet AGN  is estimated to
be in the region of  25~per cent \citep{Padovani2009}.

Both modes of AGN accretion have a postulated associated
mechanism of feedback which result in star-formation quenching
in the host galaxy. In the radiatively efficient `quasar' accretion
mode, feedback occurs as a result of quasar driven winds which
systematically remove gas from the galaxy, star-formation and
accretion activity terminate abruptly as this process depletes their
local fuel supply
\citep{Silk1998,Dimatteo2005,cattaneo}.

In the radiatively inefficient `radio' accretion mode the bulk of
the energy from the AGN is emitted as kinetic energy in jets, with
very little radiative output from a central accretion disk \citep*{Merloni2007}. These
sources accrete at much lower rates and consequently the total energy
output from these AGN is lower than in the radiatively efficient
case. However, depending on how efficiently the kinetic energy in the
jets is converted to heat in the interstellar gas, these AGN may still
have considerable potential to influence the star-formation properties
of the galaxies they inhabit as well as their larger scale
environments \citep[e.g.][]{McNamara2000,Birzan2004}. Using a scaling relationship between
radio luminosity and mechanical heating power derived from
cluster observations, \citet{Best2006} demonstrated that the
time-averaged energetic output of `radio' mode accretors was indeed
sufficient to counter cooling losses in massive, red galaxies. In this
mode the AGN is fuelled by direct accretion of the hot interstellar gas
and is also the source of gas heating. As such AGN feedback in this
mode has the potential to set up a stable feedback loop where the
accretion rate is automatically adjusted by the available supply of
hot gas \citep{Allen2006,Best2006,Hardcastle2007}.   

A key piece of evidence in determining the relative significance of
both types of AGN feedback is an accurate determination of the cosmic
evolution of the radio sources whose jets are speculated to be
enormously influential via this `radio' mode accretion. It is already
well established that powerful radio AGN undergo very rapid evolution
with their number densities increasing by a factor of $\sim$1000 out
to redshifts of $\sim$2
(\citealp{Longair1966}; \citealp*{LRL1983}; \citealp{DP1990,Willott2001}), with evidence of a decline in their number density taking place beyond 
$z{\sim}$3 \citep{JarvisRawlings00,Jarvis01b,Wall05,Rigby2011}.  While current studies
of the lower luminosity, predominantly LERG, radio population indicate
that they experience much milder positive evolution, with density
enhancements of the order 2--10 out to $z{\sim}$1.2
(\citealp{Clewley2004,Sadler}; \citealp*{Donoso}; \citealp{Smola,Strazzullo,PadovaniLF,MJ}). Recent work by
\citet{Best2012} suggests that the observed mild evolution at low
radio luminosities may be primarily driven by the evolution of a small
number low-luminosity
HERGs. They find that LERG populations experience little or no
evolution out to $z{\sim}$0.3 while
HERGs are strongly evolving at all luminosities, but only contribute a
small fraction of the low luminosity radio population.

A further division in evolutionary behaviour has been observed between
radio-quiet and radio-loud AGN.  Studies by \citet{Padovani2009} and
\citet{Simp2012} find very little evidence of evolution in the
low-luminosity radio-loud sources and evidence for stronger evolution
in the radio-quiet AGN. This possibly provides further evidence of a
link between the radio-quiet AGN and the HERG sources which are
believed to be powered by the same radiatively efficient accretion
mode, and may suggest that it is these `quasar' mode accretors which
undergo the strongest evolution.

This paper presents a new investigation of the evolution of low
luminosity radio sources out to $z{\sim}$2.5.  Details of the radio
observations and multiband photometry used in this study are presented
in sections~\ref{sec:radio} and \ref{sec:multi}. The cross matching procedure is outlined in section~\ref{sec:xmatch}. A
description of the photometric redshifts used to construct the radio
luminosity function is presented in sections~\ref{sec:photoz}. While section~\ref{sec:RLF} presents our results  and our
conclusion and discussion is presented in section~\ref{sec:conc}. We use AB magnitudes throughout this
paper and assume a cosmology of H$_{0}{=}70$~km~s$^{-1}$~Mpc$^{-3}$, $\Omega_m{=}0.3$, and $\Omega_{\Lambda}{=}0.7$.

\section[]{Radio Data}
\label{sec:radio}
This work utilises the radio survey completed by \citet{Bondi2003} using
the Very Large Array (VLA). These observations cover 1 square degree
centred at  $\alpha$(J2000)=$\mathrm{2^h26^m00^s}$ and
$\delta$(J2000)=$\mathrm{-4^{d}30' 00''}$ with 9 pointings. They were taken in the VLA
B-configuration and have a FWHM synthesised beamwidth of approximately
6~arcsec while the final mosaiced image has a background rms noise level
of $\sim$17$\mu$Jy.   
 
A catalogue of  1054 radio sources whose peak fluxes exceed 60$\mu$Jy was extracted from the mosaiced image
using the AIPS Search and Destroy (SAD) task. To minimize the
contribution of spurious sources to the final catalogue sources were
retained as real detections only if their
peak flux to local noise ratios is $>$5. Multiple component
radio sources are identified in the image by assuming their components
meet the following  three criteria, they are separated by
$<$18~arcsec, have peak flux ratios $<$3, and all components have a
peak flux $>$0.4~mJy/beam, only 19 multiple component sources are
identified in the field. Further details of the calibration,
catalogue extraction and multi-component classification procedures are
outlined in \citet{Bondi2003}.

 Further observations of this field were performed by
 \citet{Bondi2007} using the GMRT at 610~MHz. The GMRT observations
 covered the whole square degree of interest with 5 pointings observed
 for $\sim$5.5~hours each, resulting in a 5$\sigma$ limiting flux density of
 $\sim$200$\mu$Jy. These GMRT observations were used to obtain
 spectral index estimates of the VLA radio
 sources after matching within a 3'' tolerance. Throughout this paper, wherever a spectral index estimate is
 required, e.g. in the case of determining radio luminosities, we use
 the estimates from \citet{Bondi2007}. For the 269 sources not
 detected by the GMRT at 610~MHz we assume a standard spectral index of $\alpha{=}-0.7$.

\section{Multi-band photometry}
\label{sec:multi}
The square degree of VLA radio observations has been observed with the
VISTA Deep Extragalactic Observations \citep[VIDEO;][]{video} survey. The VIDEO survey
is a 12 sq. degree survey over three fields with the Visible and
Infrared Survey Telescope for Astronomy (VISTA) whose objective is to
study  the formation and evolution of galaxies and galaxy clusters
from the present day to $z{\sim}$4. The survey provides photometry in the
Z,Y,J,H and K$_s$ bands to 5$\sigma$ depths (2~arcsec apertures) of 25.7, 24.6, 24.5, 24.0,
23.5  magnitudes respectively. This field also coincides with the
Canada-France-Hawaii Telescope Legacy Survey D1 field \citep[CFHTLS D1;][]{Ilbert2006} which provides additional photometry in the u*,g',r',i',z' optical bands to depths of 26.5, 26.4, 26.1, 25.9, 25.0.  

\section{Cross-Matching}
\label{sec:xmatch}
Single component radio sources were matched to their infrared counterparts using the
Likelihood Ratio (LR) technique \citep{deruiter,Sutherland1992,Ciliegi2003}. The LR is a commonly used method 
to associate low resolution radio observations with higher resolution
optical or infrared observations.  In brief, it calculates the
ratio of the probability that a given source and counterpart are
related to the probability that they are unrelated. This probability
takes into account the positional accuracy of the near infrared and radio observations
as well as the magnitude distributions of the background infrared sources and
the radio source counterparts. Full details of the cross-matching
procedure between the VIDEO survey and the \citet{Bondi2003} VLA observations are
discussed in \citet{Mcalpine}.  

The procedure resulted in a
cross-matched catalogue of 942 radio sources whose counterparts
are brighter than  $K_{s}\sim$23.8. This represents a
completeness of 91.0~per cent, the percentage of misidentified counterparts in
this catalogue is predicted to be very low at the level of
$\sim$0.8~per cent.

Multiple component radio sources were associated with their counterparts via
visual inspection.  These sources are less suited to the LR method used
for single component sources as the likely position  of the
counterpart source and the associated errors on this position are
poorly constrained. There are 19 multiple sources in this field
and only  9 of these were associated with an infrared counterpart.

\section{Photometric Redshifts}
\label{sec:photoz}
Photometric redshifts for the combined VIDEO and CFHTLS datasets have
been determined using the publicly available code Le Phare\footnote{http://www.cfht.hawaii.edu/~arnouts/LEPHARE/lephare.html}
\citep{Ilbert2006}. This code derives photometric redshifts via a Spectral Energy Distribution (SED) fitting
method.  It operates by shifting a set of input
template SEDs, assumed to be representative of the true SED profiles
of the observed sources, to a range of redshifts and fitting these to
the observed  photometry. The redshift of the best fitting template is then adopted as the best photometric redshift
estimate. 

The accuracy of the photometric redshifts was assessed by comparing
with spectroscopic redshifts in the VIMOS VLT deep survey \citep[VVDS;][]{lefevre2005}, which is a deep 
spectroscopic survey limited to $I_{\mathrm{AB}}$ $\sim$24.0. Comparing
to sources with very secure spectroscopic redshifts and excluding
sources identified as quasars, the sample has a normalised median absolute
deviation (NMAD) in $\frac{\Delta z}{(1+z_{s})}$of $\sigma\sim$0.025, where
$\Delta z{=}|z_{p}-z_{s}|$ \citep[see][]{Ilbert2006}. Approximately
3.8~per cent of the sources are catastrophic outliers, defined
as cases where  $\frac{\Delta z}{(1+z_{s})}>$0.15. Further details of the procedure used to derive these photometric redshifts are provided in the VIDEO
survey description paper \citep{video}. 


 \subsection{Quasar photometric redshifts}

As some fraction of the radio sources will be associated with quasars
we conducted an investigation into the photometric redshifts of
quasars in the VIDEO survey.  Photometric redshifts determined for
quasars are generally much less
reliable than those obtained for galaxies due to the absence of strong
spectral break features which provide the strongest contraints in the
SED fitting procedure \citep[see e.g.][]{Richards2001,Babbedge2004,Mobasher2004,Polletta2007}. Quasar SEDs are generally well represented
by a power law continuum overlaid with a series of  broad and narrow
emission line features. As a power law spectrum is
invariant under redshift, constraints in the photometric redshift fitting procedure
rely on the  quasar emission line features. Success thus depends on
the emission line features having sufficient flux to
influence the measured broad-band photometry. These emission
lines are clearly difficult to identify and accurately localise based
solely on the final integrated flux measurements, and may also fall
between gaps in the filter coverage. Further sources of error include
the possibility of strong contamination of the AGN SED by the host
galaxy as well as the intrinsic variability of quasars. As most surveys
have non-simultaneous photometry this variability hampers the
construction of a typical snapshot SED for fitting. These factors
culminate in the effect that the most accurate photometric redshifts
for quasars, using a variety of techniques to mitigate these limitations,
produce photometric redshifts with dispersions in $\frac{\Delta
  z}{(1+z_{s})}$ of $\sim$0.35
\citep[see][]{Ball2008,Salvato2009,Salvato2011} whilst for galaxies much simpler
techniques routinely report estimates $\frac{\Delta z}{(1+z_{s})} \sim$0.1 or even much lower values \citep{Wolf2004,Ilbert2006,Ilbert2009,Cardamone2010,video}.

\begin{figure}
\includegraphics[width=1\columnwidth]{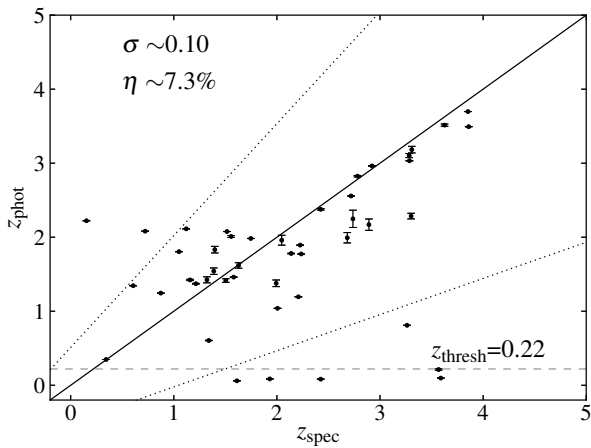}
\caption{Photometric versus spectroscopic redshifts for objects
  identified as quasars in the VVDS. The sample has a NMAD in
  $\frac{\Delta z}{(1+z_{s})}$ of $\sigma{=}$0.10 and approximately
  7.3~per cent of the
objects are 5$\sigma$ outliers. Photometric redshifts below the $z_{\mathrm{thresh}}$
level, plotted as a dashed line, are clearly not reliable. Dotted
lines are plotted at 1$\sigma$. The error bars represent the 1$\sigma$
errors derived from distribution of $\chi^2$ values as a function of z
as described in section~\ref{sec:zerr}. }
\label{fig:agnspec}
\end{figure}

To determine the severity of this reduced accuracy in photometric
redshift estimates, in figure~\ref{fig:agnspec}  we
show photometric versus spectroscopic redshifts for the 73 sources identified
as quasars in the VVDS survey. These were
classified as quasars due to the presence of broad emission line features in
their spectra. Only sources with secure redshifts were retained for comparison.  There
is a clearly discernable threshold in figure~\ref{fig:agnspec}, at
$z_{\mathrm{phot}}$=0.22, below which the photometric redshifts are
unreliable.  Reassuringly the photometric redshifts of the remaining objects
appear to be well correlated with spectroscopic redshift, although there is a larger spread in
$\frac{\Delta z}{(1+z_{s})}$ than for the general galaxy
population. This quasar sample has a NMAD in $\frac{\Delta
  z}{(1+z_{s})}$ of $\sigma {\sim}$0.10 and $\sim$7.3~per cent of the sample are catastrophic outliers with
$\frac{\Delta z}{(1+z_{s})}{>}5\sigma$. 

As figure~\ref{fig:agnspec} indicates that quasars with $z_{\mathrm{phot}}{<}$0.22 may
be unreliable, we attempted to identify such potentially
problematic sources in our sample. Candidate quasars were identified
based on their {\sc sextractor} \citep{Bertin1996}
CLASS\_STAR parameter which provides a measure of how well resolved an
object is in the infrared images. A criteria of CLASS\_STAR${>}0.8$
identified 78 potential quasars in the VLA survey and eight of these
have photometric redshifts of ${<}0.22$. To determine the effect of these potentially
poor redshift estimates on our conclusions we calculate the RLF in
section~\ref{sec:RLF} both including and excluding these 8 sources and find the
results of our analysis to be unchanged.

\subsection{Redshift Distribution}
\label{sec:zdist}

\begin{figure}
\includegraphics[width=1\columnwidth]{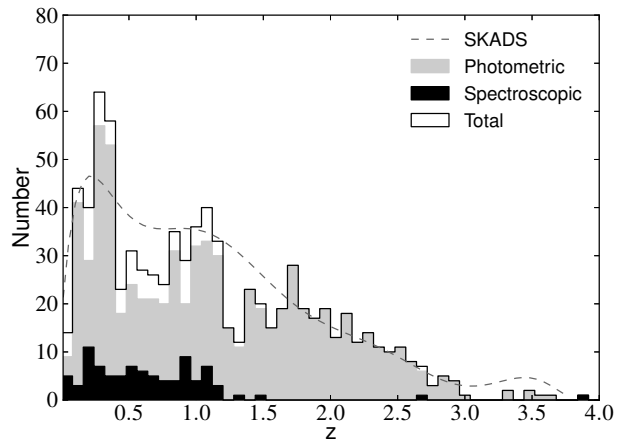}
\caption{Redshift distribution of all radio sources with counterparts
  in the VIDEO survey. Objects with spectroscopic redshifts in the
  VVDS are plotted in black, while photometric redshifts are indicated
in grey. The redshift distribution predicted by the SKA simulated
skies \citep[SKADS,][]{Wilman2008,Wilman2010} is overplotted by the dashed line.}
\label{fig:zdist}
\end{figure}

 The redshift distribution of all radio sources with counterparts in
 the VIDEO survey is presented in figure~\ref{fig:zdist}. Objects with spectroscopic
 redshifts in the VVDS survey are plotted in black. A notable feature
 of this plot is a large peak in the distribution at redshifts of
 $z\sim$0.2--0.4. 

\begin{figure}
\includegraphics[width=1\columnwidth]{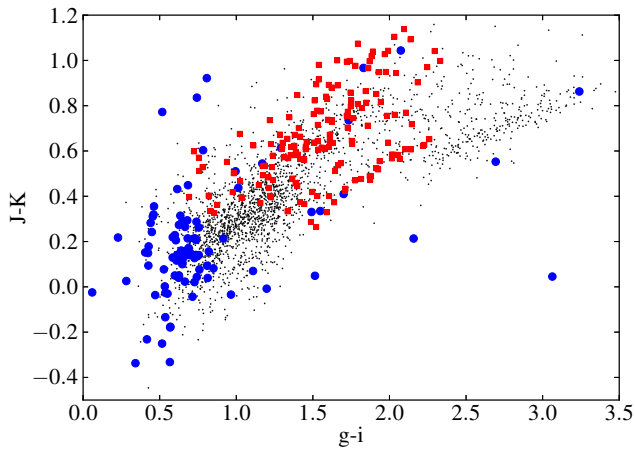}
\caption{$g$-$i$ versus $J$-$K$ colour diagram. Photometric redshift
  outliers with $\frac{\Delta z}{1+z_{\mathrm{spec}}}$ $>$0.15  are
  indicated by blue dots, while radio sources with
  0.2${<}z_{\mathrm{phot}}{<}$0.4 are plotted as red squares. The
  outlier sources preferentially occupy the bottom left hand corner of
the diagram and show little overlap with the low redshift radio
sources in their colour distribution. }
\label{fig:photozcomp}
\end{figure}

The origin of this peak is unclear as there are no
 corresponding large peaks in the redshift distribution of the full VIDEO
 infrared catalogue. Objects with
 poor photometric redshift estimates in the VIDEO survey were found to
 be preferentially associated with bluer colours
 \citep{video}. To determine whether the large peak at $z{\sim}$0.2--0.4 could be caused by a failure in the photometric redshift
 estimation process we compare the colour distribution of radio
 sources with 0.2${<}z_\mathrm{phot}{<}$0.4 to the colour distribution
 of NIR sources whose $\frac{\Delta z}{1+z_{\mathrm{spec}}}{>}$0.15.  It is clear from figure~\ref{fig:photozcomp} that the
 low redshift radio sources do not occupy the bluer region of the $g$-$i$,
$J$-$K$ colour diagram associated with poor redshift estimates, tending
 to suggest that these estimates are reasonably reliable and the
 observed peak is a real feature of the radio sources in this
 field. There is also a small peak visible in the spectroscopic
 redshifts in figure~\ref{fig:zdist} at $z{\sim}$0.2--0.4 that adds weight to our assertion that
 the photometric redshifts are reliable.

The feature at $z{\sim}$0.2--0.4 could plausibly be due to large-scale
structure within this relatively small field. We note that there are
six known X-ray clusters at $z$=0.262, 0.266, 0.293, 0.301, 0.307 and
0.345 \citep{Pacaud, Adami} in this field which are at least partially
responsible for an increase in the galaxy density within this redshift
range.  However, we also note that we expect there to be an
overdensity of radio sources around this redshift range due to the
fact that star-forming galaxies can be detected to $z{\sim}$0.4 in our
radio data, whereas beyond this redshift the radio source population
becomes dominated by AGN, as is apparent from the SKADS simulations
shown in figure~\ref{fig:zdist}.

\subsection{Photometric Redshift Errors}
\label{sec:zerr}
Comparison with spectroscopic redshifts provides an important
characterization of the variance and fraction of failures in the
photometric redshift estimates, however this process cannot be used to
identify the photometric redshifts in the full sample which are most
likely to be unreliable. Extra information regarding the reliability
of individual redshift estimates can be extracted by considering
whether the fitting procedure used to produce it was well constrained
by the available photometry or not. A measure of the uncertainty in
the photometric redshift estimate can thus be obtained directly from
the $\chi^2$ distribution obtained when fitting the template to the
observed photometry. A redshift probability density function (PDF$z$) is
constructed from the $\chi^2$ values as $\mathrm{PDF}z \propto \exp\left(-\frac{\chi^2}{2}\right)$. The 1$\sigma$ confidence intervals are
predicted by determining the redshifts corresponding to an increment
in the $\chi^2$ value of $\Delta \chi^2{=}1$, while 3$\sigma$
errors are determined at $\Delta \chi^2{=}9$. 

\begin{figure}
\includegraphics[width=1\columnwidth]{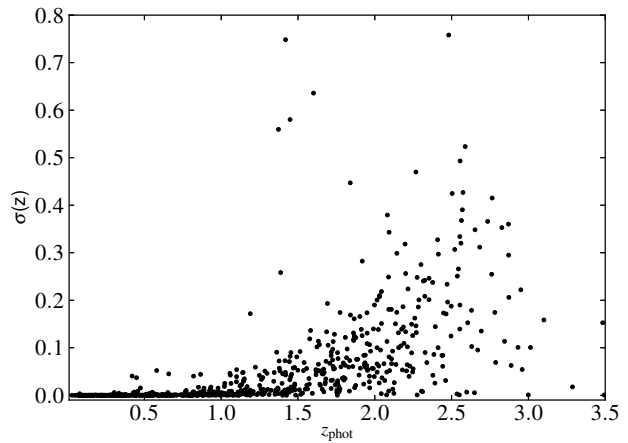}
\caption{The 1$\sigma$ errors on photometric redshift estimates as a
  function of redshift.}
\label{fig:errors}
\end{figure}

Errors produced by this method do not account for intrinsic
uncertainties in the photometry not included in the measured
photometric errors such as those caused by blending or the presence of
bright neighbours, nor can they account for inadequacies in the input
template library. They are nevertheless useful indicators of
reliability in the absence of spectroscopic data and
\citet{Ilbert2006} demonstrated that 68~per cent and 92~per cent of the
spectroscopic redshifts are located within the 1$\sigma$ and 3$\sigma$
errors respectively. Figure~\ref{fig:errors} presents the errors estimated from the
$\chi^2$ fitting for the radio sources in this sample. This figure
demonstrates that the uncertainty in the fitting
procedure, and consequently the redshift estimates,  increases
significantly towards
higher redshifts. This is to be expected due to the larger errors on
faint photometry towards high redshift objects as well as the possibility
that the locally observed, empirical templates used in the SED fitting procedure
are not sufficiently representative of these high redshift
objects. These larger errors towards higher redshifts are accounted for
in our determination of the Radio Luminosity Function (RLF) via
Monte Carlo simulations which account for the probability distribution
in $z_{\mathrm{phot}}$ for each radio source.

\section{Radio Luminosity Function}
\label{sec:RLF}
We determine the RLF for the full population of
radio sources in six redshift bins using the standard $\frac{1}{V_{\mathrm{max}}}$
method \citep{Schmidt}. Inclusion in the final cross-matched sample depends on the
source exceeding the flux limits of both the radio and 
near-infrared surveys, thus the final
maximum observable redshift $z_{\mathrm{max}}$ is calculated as min($z_{\mathrm{radio}}$,
$z_{\mathrm{IR}}$) where $z_{\mathrm{radio}}$ and $z_{\mathrm{IR}}$
represent the maximum observable redshifts of the source in the radio and infrared
surveys respectively. We estimate the maximum $z_{\mathrm{IR}}$ by
redshifting the best fitting SED template from the photometric redshift
estimation procedure and  determine the redshift where the template becomes
fainter than our imposed magnitude limit of K$_{s}$=23.8. Similarly
for the radio sources we estimate their intrinsic luminosity assuming
the standard spectral index based k-correction and determine the
redshift at which they drop below the VLA survey flux limit. 
 
\begin{figure}
  \includegraphics[width=1.0\columnwidth]{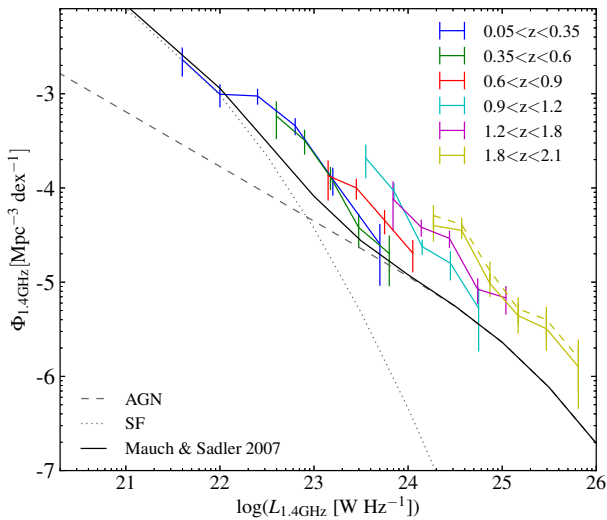}
\caption{Radio luminosity function in six redshift bins to
  $z{\sim}$2.5. The error  bars incorporate the uncertainty in
  photometric redshifts, estimated
  via Monte Carlo simulations. The dashed yellow line represents the
  change in the RLF which occurs at the highest redshifts
  assuming that the unmatched radio sources in the VIDEO survey have a
redshift distribution similar to the faintest sources, $K_{s}{>}23.5$, in the
SXDF \citep{Simp2012} field. The solid black line represents the local
luminosity function of \citet{MauchSadler2007} with the contribution
of  star-forming
galaxies and AGN represented by the dotted and dashed lines
respectively.}
\label{fig:rlfall}
\end{figure}

Figure~\ref{fig:rlfall} presents the RLF in six redshift bins in the interval $z$=0--2.5,
the error bars in the plot incorporate both the errors due to the
sample size per bin as well as the uncertainties in
the photometric redshift estimates.  The latter were estimated using
Monte Carlo simulations by constructing 1000 possible
realisations of the PDF$z$  of the photometric redshifts,
calculating the RLF for each of the realisations and determining the
median and standard deviation of these simulated RLFs. Including
the effects of uncertainty in the photometric redshifts is
particularly relevant in the high redshift bins as
figure~\ref{fig:errors} indicates that these errors increase
substantially at $z{>}$1.4. In order to account for sample
variance in a field of 1~square degree we determine the variance in
the number of sources ${>}100~\mu$Jy in 100 1~square degree fields in the SKADS
simulations (\citealp{Wilman2008,Wilman2010};\citealp*[see also][]{HJC}). This variance was
determined for each of the redshift bins considered in figure~\ref{fig:rlfall}, and added in quadrature to the errors due to photometric
redshift uncertainties. The calculated variances are reported in
table~\ref{fig:varskads}.

\begin{table}

\centering
\caption{The variance on the number of radio sources above 100~$\mu$Jy in a
  1~square degree field as function of redshift, determined from the
  SKADS simulations \citep{Wilman2008,Wilman2010}}
\label{fig:varskads}
\begin{tabular}{ll}
$z$ & $\sigma$ [per cent]  \\
\hline 
0.1--0.35& 15\\
0.35--0.6& 13\\
0.6--0.9 & 15\\
0.9--1.2 & 18\\
1.2--1.8 & 14 \\
1.8--2.5 & 14

\end{tabular}
\end{table}

The RLF in this work is compared to the local RLF determined by
\citet{MauchSadler2007} using NVSS radio sources with spectroscopic
redshifts from the 6 degree Field Galaxy Redshift Survey (6dFGRS;
\citealt{Jones2004}). The \citet{MauchSadler2007} luminosity function
is constructed from separate fits to the luminosity functions of
star-forming galaxies and AGN, and the individual contributions
of these two populations are plotted as dotted and dashed lines in
figure \ref{fig:rlfall} respectively. It is clear that the
local radio luminosity function is dominated by star-forming galaxies and AGN
contributions at low and high luminosities respectively, with the
division falling at roughly $\sim10^{23}$~$\mathrm{W~Hz^{-1}}$. There
is a large discrepancy between the VLA-VIDEO local RLF determined in
this work and that of the \citet{MauchSadler2007} star-forming galaxy
luminosity function at low luminosities (reduced $\chi^2$=5.9). This discrepancy is probably
caused by the large concentration of objects in the VLA-VIDEO survey
in the range $z{\sim}$0.2--0.4 (see figure~\ref{fig:zdist}), which can easily be explained by sample variance at these low redshifts, where our volume is relatively small. The
absence of higher luminosity radio sources in the lowest redshift bin,
due to the small sky area, precludes us from making a direct comparison with the local AGN
luminosity function.

\subsection{Evolution of combined star-forming and AGN population}
\label{sec:onecomp}
Figure \ref{fig:rlfall} appears consistent with a fairly rapid
increase in the space density of radio galaxies up to
redshifts of $z\sim$ 1.2, with hints of a possible slowing down of this
process between
redshifts $z=1.2$--2.5. To quantify these broad evolutionary trends we initially
use models of both pure density and pure luminosity evolution of the
combined star-forming and AGN RLF, such
that:\begin{equation}\Phi_{\mathrm{PDE}}(z)=(1+z)^{\alpha_{d}}(\Phi_{\rm{SF}}+\Phi_{\rm{AGN}}) \end{equation}
\begin{equation}
\Phi_{\mathrm{PLE}}(z)=\Phi_{\rm{SF}}\left(\frac{L}{(1+z)^{\alpha_{l}}}\right) +\Phi_{\rm{AGN}}\left(\frac{L}{(1+z)^{\alpha_{l}}}\right)   \\                                                                                                                                                                                                                                                             \end{equation}
where $\Phi_{\rm{SF}}$ and $\Phi_{\rm{AGN}}$ are the star-forming and
AGN luminosity functions of \citet{MauchSadler2007}
respectively. $\Phi_{\mathrm{PDE}}(z)$ and $\Phi_{\mathrm{PLE}}(z)$
are the evolved luminosity functions at redshift $z$ assuming pure
density and pure luminosity evolution respectively. Such
models are clearly limited in their ability to accurately represent the
behaviour of the three sub-classes of potentially independently
evolving radio sources present in the sub-mJy VLA-VIDEO sample. The
parametrization simply provides a convenient means to quantify the
overall strength and sense of the evolution taking place at different
cosmic times. 

Based on the tentative evidence in figure \ref{fig:rlfall} that the
evolutionary behaviour of these sources changes at a redshift of
$\sim$1.2, we fitted six possible evolutionary scenarios to the
data, the first is a pure density evolution model with a single
$\alpha_d$ parameter for the entire redshift range from
$z{\sim}$0--2.5. The second is of pure density evolution out to a
redshift of $z$=1.0 and no further evolution taking place beyond this,
and the third fits independent pure density evolution parameters
$\alpha_{d1}$ and $\alpha_{d2}$ in the redshift $z{\sim}$0--1.2 and
1.2--2.5 ranges. We repeat the fitting procedure for these three
scenarios assuming pure luminosity evolution. To avoid the fit being
biased by the large discrepancy between the VLA-VIDEO RLF and the \citet{MauchSadler2007} RLF at
low redshifts we excluded the $z{<}$0.35 points from the fit. 

\begin{figure*}
\includegraphics[width=1.85\columnwidth]{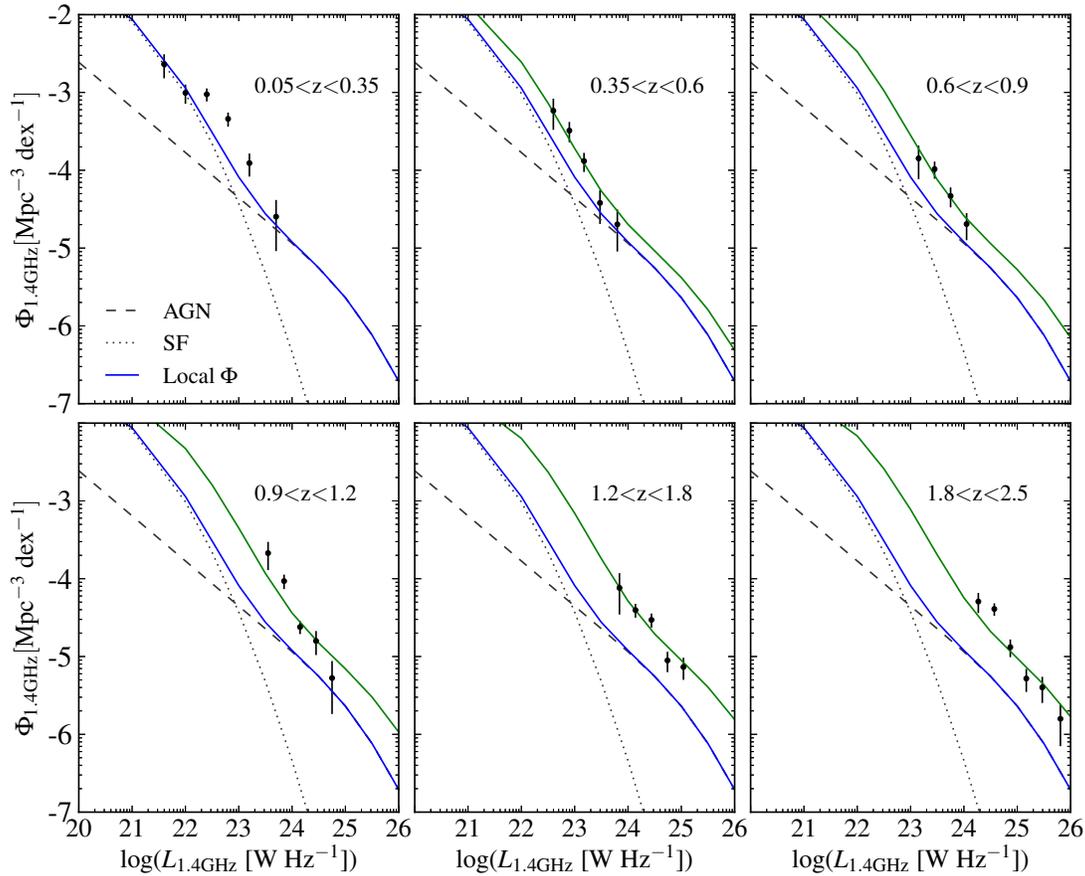}
\caption{The radio luminosity function in six redshift bins to
  $z{\sim}$2.5. The local luminosity function from
  \citet{MauchSadler2007} is plotted as a blue solid line. The
  contributions from star-forming galaxies and AGN
  to the local RLF are plotted as black dotted and dashed lines respectively. The green solid line represents the evolved luminosity
  function of the best fitting evolutionary scenario. Namely pure
  luminosity evolution with $\alpha_{l1}{=}1.90$ and $\alpha_{l2}{=}0.26$.}
\label{fig:rlfmany}
\end{figure*}

The evolution parameters
and $\chi^2$ values determined from these fits are presented in table
\ref{tab:evol}, and figure~\ref{fig:rlfmany} presents a comparison of
the data points to the evolved luminosity function produced by the
best fitting model. In all cases the data implies an increasing source
density out to $z{\sim}$1, with density enhancements at this redshift of
a factor of $\sim$3 over the local values. The independent
$\alpha_{d2}$ and $\alpha_{l2}$ determined in model 3 imply an increase
in space density towards earlier cosmic times but are also consistent
with the scenario of no evolution beyond $z{\sim}$1.2 (within 2$\sigma$). A constant positive evolution across the
entire redshift range provides a better fit to the data in the pure
density evolution case, while a slower evolution beyond $z{\sim}$1.2 is
a better fit for pure luminosity evolution.

\begin{table}
\centering
\caption[$\chi^2$ fits to radio luminosity function]{Results of fitting three different evolutionary scenarios to the RLF in six redshift bins. Pure density evolution is assumed in all three cases.}

\newcommand{\mc}[3]{\multicolumn{#1}{#2}{#3}}
\begin{tabular}{lllllc}
&× & z=0--1.2 & z=1.2--2.5 & ×&\\&× & $\alpha_{d1}$ & $\alpha_{d2}$ & $\chi^2$ & reduced $\chi^2$\\
PDE&model 1 & \mc{2}{l}{1.28$\pm$0.88}& 34.61 & 1.44\\
PDE&model 2 & 1.53$\pm$0.11 & 0$\pm$0  & 36.48 & 1.52\\
PDE&model 3  & 1.38$\pm$0.17& 0.23$\pm$0.16 & 33.63 & 1.46

\end{tabular}

\label{tab:evol}
\end{table}

\begin{table}
\centering
\caption[$\chi^2$ fits of pure luminosity evolution]{Results of fitting three different evolutionary scenarios to the RLF in six redshift bins. Pure luminosity evolution is assumed in all three cases.}

\newcommand{\mc}[3]{\multicolumn{#1}{#2}{#3}}
\begin{tabular}{lllllc}
&× & z=0--1.2 & z=1.2--2.5 & ×&\\
&× & $\alpha_{l1}$ & $\alpha_{l2}$ & $\chi^2$ & reduced $\chi^2$\\
PLE&model 1 & \mc{2}{l}{1.72$\pm$0.11}& 34.67 & 1.45\\
PLE&model 2 & 2.03$\pm$0.13 & 0$\pm$0  & 33.03 & 1.37\\
PLE&model 3  & 1.90$\pm$0.17 & 0.26$\pm$0.19 & 30.66 & 1.33

\end{tabular}
\label{tab:evol2}
\end{table}


Recent results from \cite{Simp2012} identified tentative evidence of a
decline in the RLF of radio sources at redshifts $>$1.5 in the
luminosity range 10$^{24-25.5}$~W~Hz$^{-1}$ (see their figure 11),
however the VLA-VIDEO RLF consistently increases with redshifts even
beyond $z{\sim}$1.  \citet{Rigby2011} also identify a luminosity dependent
turnover in the luminosity function, with lower luminosity objects
experiencing a decline in their number densities at lower redshifts
than their high luminosity counterparts. Their results imply a
turnover at $z{>}$0.7 for objects with luminosities in the
10$^{25-26}$~W~Hz$^{-1}$ range, which is at a slightly lower redshift
than seen in the VIDEO-VLA RLF. The points in figure~\ref{fig:rlfall}
do not betray any hint of such a luminosity dependent effect. However
the use of a single flux-density limited sample restricts our
investigation to a
narrow luminosity range in each redshift bin hampering any attempt to confirm this.

\subsection{Evolution of separate star-forming and AGN populations} 
\begin{figure*}
\includegraphics[width=1.85\columnwidth]{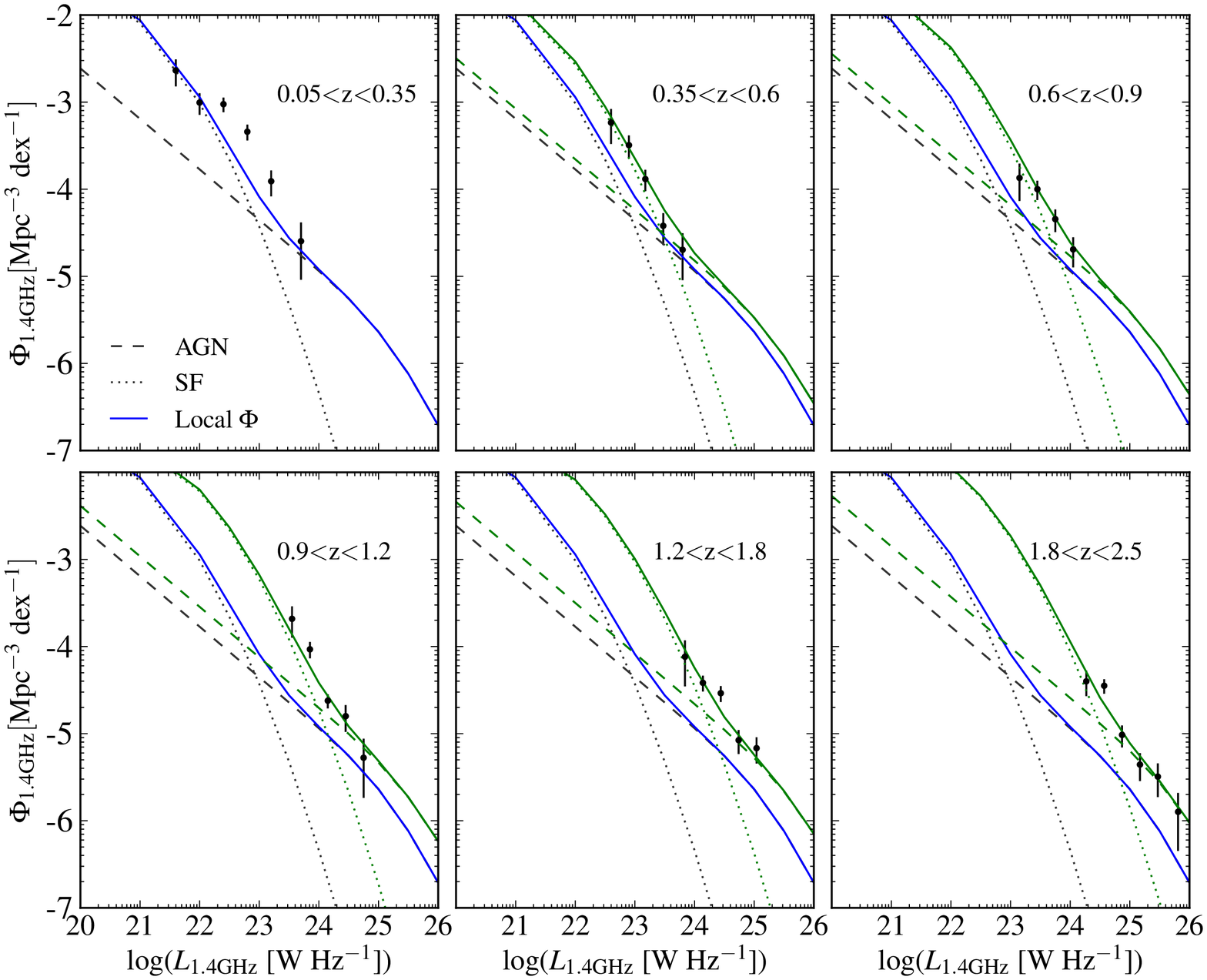}
\caption{Radio luminosity function in six redshift bins to
  $z{\sim}$2.5. The local luminosity function from
  \citet{MauchSadler2007} is plotted as a blue solid line. The
  contributions from star-forming galaxies and AGN
  to the local RLF are plotted as black dotted and dashed lines respectively.The green solid line
  represents the best fit to the data when allowing the AGN and
  star-forming RFL to evolve independently. The best fit model is pure
  luminosity evolution with $\alpha_{l}^{\mathrm{AGN}}{=}1.18$ and
  $\alpha_{l}^{\mathrm{SF}}{=}2.47$. The green dotted line is the
  evolved RLF of star-forming galaxies and the green dashed line is
  the evolved RLF of AGN. }
\label{fig:rlfmanysep}
\end{figure*}

AGN and star-forming populations may evolve independently from one another and there
is some evidence that low luminosity AGN evolve
more slowly than star-forming galaxies up to $z{\sim}$1.2 \citep[e.g.][]{Smola,Smolb}.  Figure 5 indicates
that at low redshifts we are primarily probing luminosities dominated by
star-forming galaxies, whereas at higher redshifts the the radio source
population is dominated by contributions from  AGN. Thus the observed
decline in the evolution of the LF towards higher redshifts ($z{>}$1.2) could be explained if the  AGN population evolved more slowly
than the star-forming galaxies in the $z{<}$1.2 interval.  To
further investigate the separate contributions of  star-forming galaxies and AGN
to the evolution of the total RLF we fit    
a two component model of pure density and pure luminosity
evolution. This model allowed the star-forming and AGN populations to
evolve independently such
that: \begin{equation}\Phi_{\mathrm{PDE}}(z)=(1+z)^{\alpha_{d}^{\mathrm{AGN}}}\Phi_{\mathrm{AGN}}+(1+z)^{\alpha_{d}^{\mathrm{SF}}}\Phi_{\mathrm{SF}} 
\end{equation}
\begin{equation}
\Phi_{\mathrm{PLE}}(z)=\Phi_{\rm{SF}}\left(\frac{L}{(1+z)^{\alpha_{l}^{\mathrm{SF}}}}\right)
+\Phi_{\rm{AGN}}\left(\frac{L}{(1+z)^{\alpha_{l}^{\mathrm{AGN}}}}\right)
\\                                                                                                                                                                                                                                                             \end{equation}
 Fitting separate  $\alpha_{l}$ and $\alpha_{d}$ values for AGN and
star-forming galaxies resulted in much lower reduced $\chi^2$ values than the single
component model in section~\ref{sec:onecomp}. The AGN population was found to evolve more slowly than the
star-forming population, thus confirming that the AGN are responsible for the
slower evolution beyond $z{\sim}$1.2 detected in the single component
fit. The results of the fit are presented in table~\ref{tab:twocomp}
 and figure~\ref{fig:rlfmanysep} compares the best fit two component model  with the
 measured  RLF.

\begin{table}
\centering
\label{tab:twocomp}
\caption{Results of fitting independently evolving star-forming and AGN
  luminosity functions to the RLF, both pure density evolution (PDE)
  and pure luminosity evolution (PLE) are considered.}
\begin{tabular}{llllc}
& \multicolumn{1}{c}{SF} & \multicolumn{1}{c}{AGN} &  & \\
 & $\alpha^{SF}$ & $\alpha^{AGN}$ & $\chi^2$ & reduced $\chi^2$
\\

PDE & 3.50$\pm$0.40 & 1.16$\pm$0.09 & 24.52 & 1.06\\
PLE & 2.47$\pm$0.12 & 1.18$\pm$0.21 &  20.64 & 0.89

\end{tabular}
\end{table}

\subsubsection{Template Classification}

\begin{figure*}
\includegraphics[width=1.85\columnwidth]{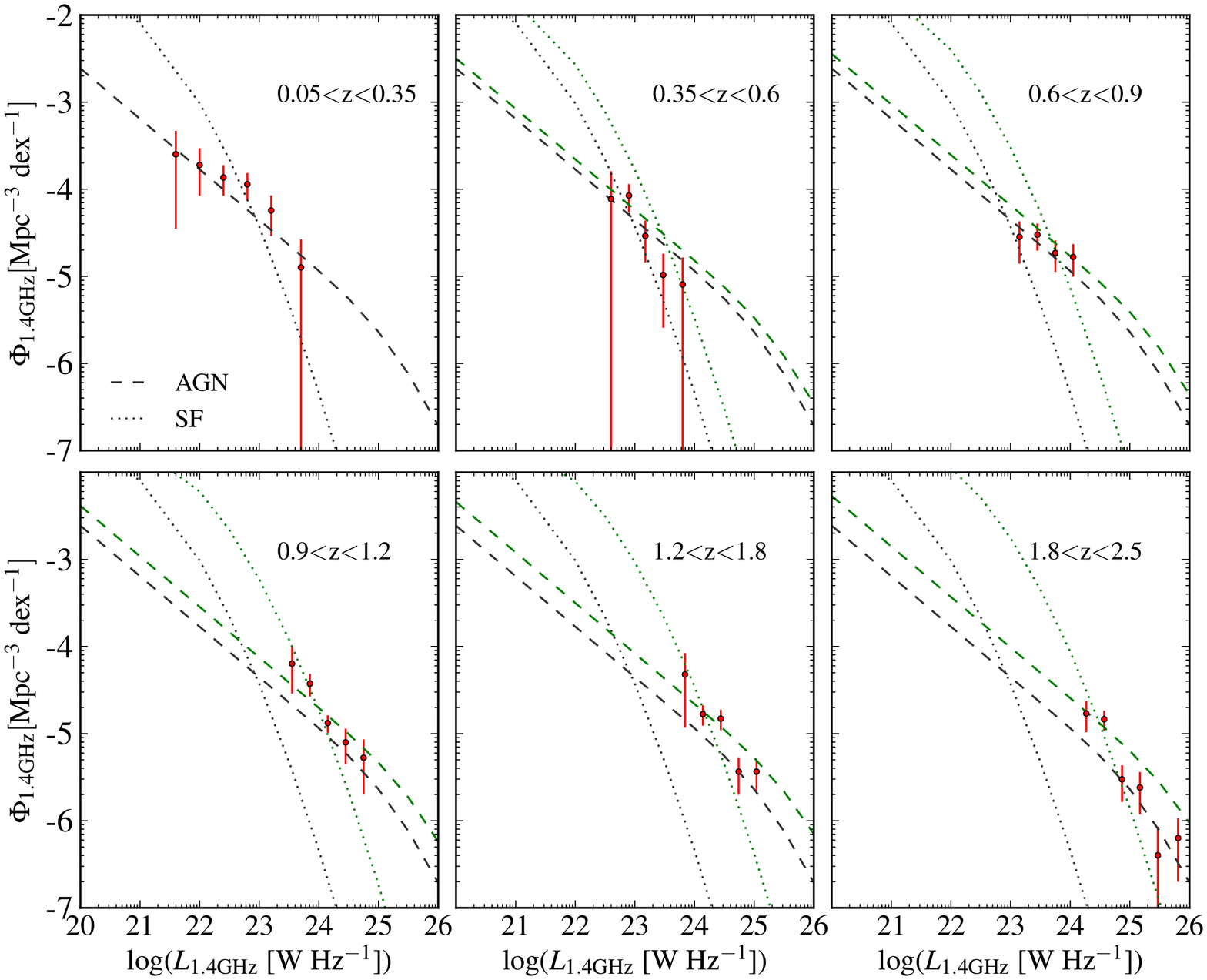}
\caption{AGN radio luminosity function in six redshift bins to
  $z{\sim}$2.5, plotted as red points. AGN are identified as objects fitted by red templates in
  the photometric redshift fitting procedure. The local luminosity
  function for star-forming galaxies and AGN \citep{MauchSadler2007} are plotted as black dotted and dashed lines respectively. The green dotted line is the
  evolved RLF of star-forming galaxies and the green dashed line is
  the evolved RLF of AGN assuming the evolution parameters in table~\ref{tab:twocomp}. }
\label{fig:agncolor}
\end{figure*}

\begin{figure*}
\includegraphics[width=1.85\columnwidth]{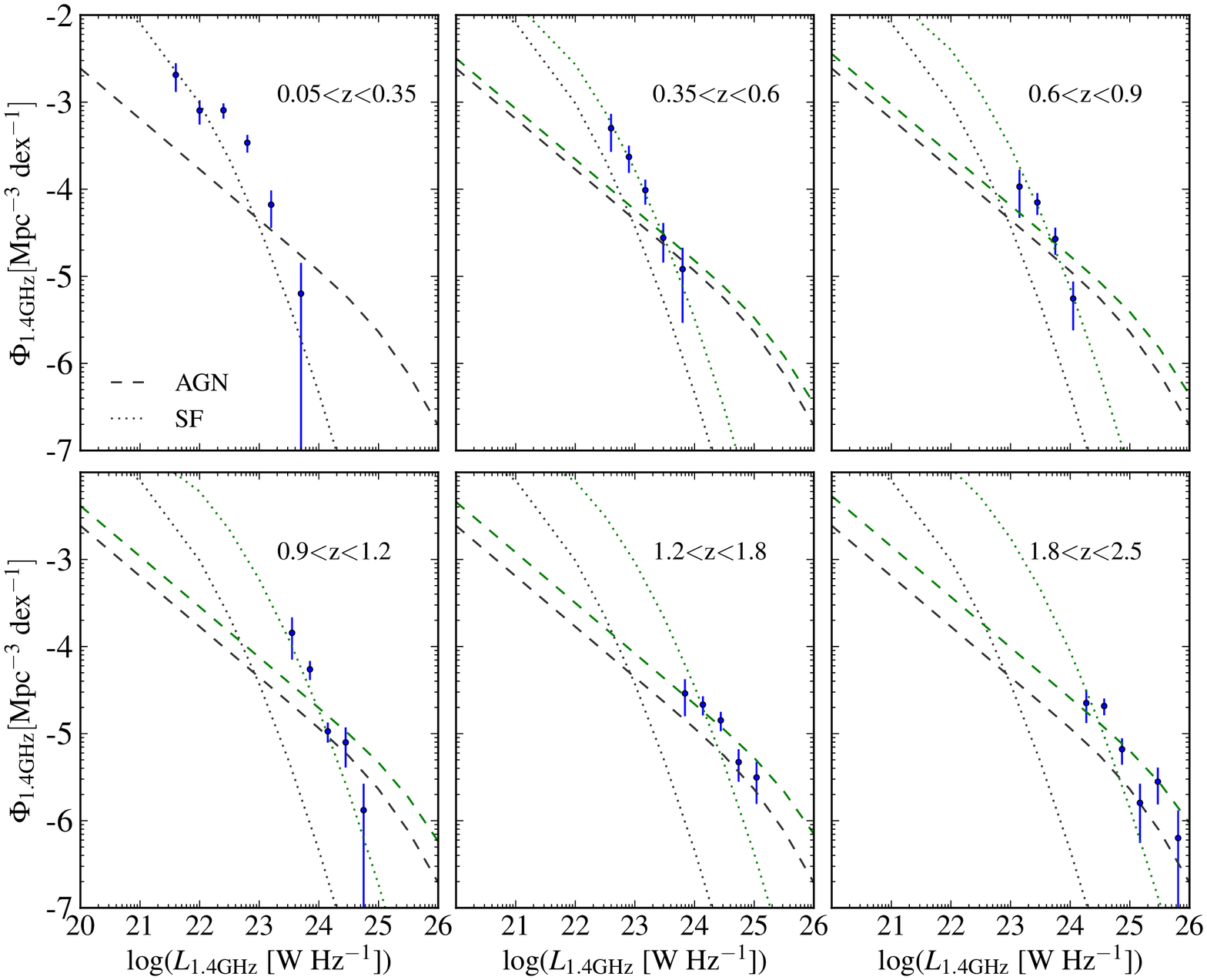}
\caption{Star-forming radio luminosity function in six redshift bins to
  $z{\sim}$2.5, plotted as blue points. Star-forming galaxies are identified as objects fitted by blue templates in
  the photometric redshift fitting procedure. The local luminosity
  function for star-forming galaxies and AGN \citep{MauchSadler2007} are plotted as black dotted and dashed lines respectively. The green dotted line is the
  evolved RLF of star-forming galaxies and the green dashed line is
  the evolved RLF of AGN assuming the evolution parameters in table~\ref{tab:twocomp}. }
\label{fig:sfcolor}
\end{figure*}

The template fitting used in section~\ref{sec:photoz}  to derive
photometric redshifts can be used to distinguish  galaxies which are star-forming
from those with predominantly old stellar populations.  As the low
luminosity radio AGN population is dominated by LERGs \citep{Best2012}
which are preferentially hosted in red passively evolving
galaxies \citep{Baldi2008,Herbert2010,Smol2011,Best2012} a template based classification scheme can be used to roughly
separate the radio population into its AGN and star-forming
components. The photometric redshift fitting procedure is based on 
templates interpolated from six observed spectra an elliptical, two spirals and and irregular galaxy as well as
two observed starburst templates. We identified AGN in the radio population
as sources redder than the spiral galaxy templates.  The
resulting AGN and star-forming RLF's, shown in figures~\ref{fig:agncolor}
and \ref{fig:sfcolor}, were found to be consistent with the fit presented in figure~\ref{fig:rlfmanysep} for these two populations out to $z{\sim}1.2$.
At higher redshifts this agreement breaks down and a large fraction
of the AGN at $z{>}1.2$ are associated with bluer star-forming
templates.

 There are several possible reasons for this, fainter photometry and greater uncertainty in the
level of dust-extinction towards high redshift objects make it harder to accurately constrain
their intrinsic underlying SEDs, thus the colour classification at these redshifts may no
longer be reliable. A second possibility is raised by the results of \citet{Jan2012} who
demonstrate that in the local universe a sub-population of LERGs are
hosted in blue star-forming galaxies,
with these blue LERGs becoming increasingly important at higher radio powers. Thus it is
possible that the contribution of such blue LERGs increases towards higher redshifts, rendering
the initial assumption that all AGN are hosted by red passive galaxies
invalid. Furthermore towards higher redshift we are probing towards
higher luminosities, increasing the contribution of HERGs which are
also preferentially hosted in galaxies with bluer colours \citep{Best2012}. 

\subsection{Unmatched radio sources}
We considered whether excluding the 103 radio sources without
reliably identified counterparts in the VIDEO survey is likely to
significantly alter the RLF calculated in this paper. We achieve this by 
making use of  redshift information available for
radio sources in the Suburu/{\em XMM Newton} Deep
field (SXDF). \citet{Simp2012} have acquired complete redshift information
(505/509 sources have redshifts) for this field by combining
spectroscopic and photometric redshifts from existing deep
optical and infrared photometry from Suburu \citep{Furusawa}
and  the UKIRT Infrared Deep Sky Survey \citep[UKIDSS;][]{Law} Ultra
Deep Survey (UDS). As the unmatched counterparts to the VIDEO
VLA survey are likely to be fainter than the adopted limits of $K_{s}$=23.8,  we  made
the simplifying assumption that their redshift distribution could be approximated
by that of the faintest, $K_{s}{>}$23.5, radio sources in the SXDF
field. This is unlikely to be an exact match for the redshift
distribution of the unmatched VIDEO sources which appear to be
fainter than the limiting magnitude of the UDS survey used in
\citet{Simp2012}, but is the most complete redshift information
available at present. 

The RLF calculated by including these non-identified
radio sources via Monte Carlo simulations was found to be consistent
with the RLF presented in figure~\ref{fig:rlfall} in all but the
highest redshift bin. At the highest redshifts the luminosity function
increases slightly when incorporating the unmatched sources, this
slight increase is indicated by the dashed yellow line in
figure~\ref{fig:rlfall}.  Thus including the unmatched sources results
in stronger
evolution being measured for AGN in the two component fit, and
stronger evolution at $z{>}1.2$ for the combined RLF in the single component fit. The fitted values
remain consistent with those reported in table~\ref{tab:twocomp} within the errors. 
 

\section{Conclusions}
\label{sec:conc}
This paper has presented a new determination of the RLF for the VLA-VIDEO
survey field using reliably identified sources with $\sim$10 band
photometric redshifts out to z $\sim$2.5. The luminosity function
broadly implies an increase in the space density of low luminosity ($< 10^{26}$~W~Hz$^{-1}$)
radio sources by a factor of $\sim$3 in the z=0--1.2 range and
is consistent with slightly slower evolution out to
$z\sim$2.5. Star forming galaxies appear to drive the stronger
evolution at $z{<}1.2$,  while at higher redshifts low
luminosity AGN dominate the sample, and their relatively weaker
evolution results in the observed slowing down of the
evolution of the RLF at $1.2{<}z{<}2.5$.

\begin{table*}
\begin{minipage}{14.1cm}
\caption[Evolution of the RLF]{Comparison of the current
  determinations of the evolution of the radio luminosity function out to a redshift of $\sim$1.3. Numbers quoted are the $\alpha_L$ parameters determined by fitting pure luminosity evolution. }
\label{tab:rlfcomp}
{
\small
\newcommand{\mc}[3]{\multicolumn{#1}{#2}{#3}}
\begin{center}
\begin{tabular}{lllllll}
Reference & \underline{Field} & \mc{2}{c}{AGN} & \mc{2}{c}{Star-forming} & All\\
× & × & {\small R-Quiet} & {\small R-Loud} & {\small Strong}
& {\small Intermediate} & ×\\\hline
& \mc{1}{l|}{}& \mc{2}{c|}{}&\mc{2}{c|}{}& \\
{\small \citet{Strazzullo}} & \mc{1}{l|}{\small SWIRE} & \mc{2}{c|}{2.7$\pm$0.3} & 3.2$^{+0.4}_{-0.2}$ & \mc{1}{c|}{3.7$^{ +0.3}_{-0.4}$ }& 3.5$\pm$0.2\\
{\small \citet{Smola,Smolb}} &\mc{1}{l|}{\small COSMOS} & \mc{2}{c|}{0.8$\pm$0.1} & \mc{2}{c|}{2.1$\pm$0.2} & \ldots\\
{\small \citet{PadovaniLF}} & \mc{1}{l|}{\small {\em Chandra} DFS} &3.8$^{+0.7}_{-0.9}$ & \mc{1}{c|}\ldots& \mc{2}{c|}{3.1$^{+0.8}_{-1.0}$} & \ldots\\
{\small  \citet{Simp2012}}& \mc{1}{l|}{\small SXDF}& $>0$ &
\mc{1}{c|}{0 \footnote{L$_{\rm 1.4GHz}\leq 10^{24}$ W Hz$^{-1}$}
}&\ldots &\mc{1}{c|}{\ldots} & $>0$\\
{\small This work}& \mc{1}{l|}{\small VIDEO} & \mc{2}{c|}{1.18$\pm$0.21}&
\mc{2}{c|}{2.47$\pm$0.12}& 1.90$\pm$ 0.17

\end{tabular}
\end{center}
}%
\vspace{-7pt}\renewcommand{\footnoterule}{}
\end{minipage}
\end{table*}

Interpreting and comparing these results to previous studies is
complicated as the radio population at these lower flux densities
contains contributions from radio-loud and radio-quiet AGN  as well as
star-forming galaxies, and these populations may all evolve
independently. Previous investigations by
\citet{Clewley2004},\citet{Sadler} and \citet{Donoso} of large samples of bright NVSS and
FIRST radio samples, with flux limits of a few mJy, imply that the low
luminosity population (${<}10^{25}$~W~Hz$^{-1}$) increases by a factor
of $\sim$2 in the z=0--0.55 range. These studies also found evidence
that the strength of the evolution taking place increases towards
higher luminosities. Further support of a luminosity dependent
behaviour was presented in \citet{Rigby2008} who found slightly stronger
positive evolution for a smaller, fainter sample of FRI objects to redshift $\sim$1. Their results imply density enhancements of $\sim$5--9 for sources brighter than the $10^{25}$~W~Hz$^{-1}$ threshold. As this work used morphologically 
identified FRI candidates, their sample is free from contamination from
star-forming galaxies and these three studies should be primarily
sensitive to evolution taking place in radio-loud AGN. The evolution
found for AGN in the VIDEO field is slightly less 
than found by \citet{Sadler} and \citet{Rigby2008}, but similar to levels
detected by \citet{Donoso}, however these works all probe the AGN
evolution at low redshifts where the constraints on the AGN in our
study are primarily at $0.9{<}z{<}2.5$. 

There have also been several characterisations of the evolution of radio sources using very deep surveys over smaller fields, these utilise a variety of criteria to separate contributions from the different underlying populations present at fainter flux densities. Their results are thus affected by uncertainties in the completeness and contamination produced by the specific classification method employed in each case, and the use of different classification methods on a per study basis also complicates attempts to make direct comparisons between them. A summary of the results of these deeper narrower studies, which probe the luminosity function over a similar redshift range as the VIDEO study in this thesis, is presented in table~\ref{tab:rlfcomp}. These works include the \citet{Smola,Smolb} studies of the COSMOS field which use a colour based separation criteria developed from the BPT diagram to identify AGN and star-forming galaxies. Their results found pure luminosity evolution with $L^*\propto (1+z)^{0.8}
$ for AGN and slightly stronger evolution in star-forming galaxies
with $\alpha_{L}\sim$2.1, with no separate classification for
radio-quiet AGN.  Their results for both AGN and star-forming galaxies thus
agree reasonably well with ours, with only slightly less evolution
found for
the AGN population.

In the SWIRE survey \citet{Strazzullo} employed a method based on SED
template fitting to separate their very faint radio sample (5$\sigma {\sim}
14\mu$Jy) into strong and intermediate star-forming galaxies and
AGN and found similar pure luminosity evolution parameters of
$\alpha_L{\sim}$3.0 for all three of these populations. This is 
higher for the AGN population than implied by the results of
\citet{Sadler} and \citet{Smola} and in the work presented here.  However it is consistent with previous estimates of the evolution taking place
in radio-selected star-forming populations and compares well with our
reported $\alpha_l$ of 2.47. \citet{PadovaniLF}  identified  pure
luminosity evolution  at levels comparable to the SWIRE survey results taking place in both their star-forming and radio-quiet quasar samples. However they find evidence that the low luminosity radio-loud AGN population undergoes no evolution in the redshift range probed by the \citet{Smola} study and suggested the evolution detected for low-luminosity 
AGN in the COSMOS 
study is driven by radio-quiet AGN included by their selection
criteria. The evolution found for AGN sources in the VIDEO survey
falls well below that found in the \citet{PadovaniLF} radio-quiet population, as would be expected if the radio quiet sources form a part of the strongly evolving `quasar' mode sources.

\citet{Simp2012} see evidence of an increase in the number of low
luminosity sources  up to $z{\sim}$1, their RLF increases
by a factor of  $\sim$3 which is consistent with our work over this redshift range. As was the case in
\citet{PadovaniLF}, they identify
this enhancement as being predominantly driven by evolution of the radio-quiet
objects. Radio quiet objects in their sample are identified based the
ratio of mid-infrared (24 $\mu m$) to radio flux and this definition
encompasses both star-forming and radio-quiet AGN. Whereas they see no
evidence of evolution for low-luminosity radio-loud AGN.  

The evolution of
  star-forming galaxies has also been extensively studied using
  optical and infrared surveys.  Mid and far-infrared {\em Spitzer}
  observations imply a galaxy population undergoing pure
 luminosity evolution with $\alpha_{l}{\sim}$3.4--3.8 out to
  $z{\sim}$1.2
  \citep[e.g.][]{Caputi2007,Magnelli2009,Ruj2010,Magnelli2011}. While
  far-infrared luminosity functions from {\em Herschel} data result in
  slightly stronger evolution estimates with $L_* \propto (1+z)^{4.1\pm0.3}$
  up to $z{\sim1.5}$ \citep{Grup2010,Lapi}. At lower redshifts  ($z{<}0.5$) some {\em
    Herschel} studies have seen evidence of stronger evolution in
  star-forming galaxies with the total luminosity density evolving as
  $(1+z)^{7.1}$ \citep{Dye2010}.  These recent infrared studies are broadly consistent with the radio derived
  star-formation histories in table~\ref{tab:rlfcomp}  implying a very similar
  increase in the total cosmic star formation density in the interval
  between $z{\sim}$0--1.  The radio results are also in good agreement
  \citep[e.g.][figure~6]{Smolb}
   with
  earlier works using UV-derived star-formation estimates 
  (\citealp{Wolf2003, Arnouts2005,Baldry2005}, \citealp[see also][]{HB2006}).


Beyond $z\sim$1.2 there have been some suggestions of a
cutoff in the RLF \citep{Waddington,Rigby2008,Rigby2011}. The LF of \citet{Simp2012} also implies
systematically lower space densities for low radio luminosity objects at redshifts $>$1.5
whilst \citet{PadovaniLF} claim evidence of strong negative evolution
in the more powerful radio-loud populations which they suggest might
be the result of a high redshift cutoff in these source
populations. However figure~\ref{fig:rlfall} indicates that the RLF
continues to increase out to $z{\sim}$2.5, although the rate of increase
does slow beyond $z{\sim}$1.2. Disentangling the contributions of the various low-luminosity radio  populations is clearly a difficult proposition and has important implications for our ability to deduce their evolutionary behaviour and make inferences about links between AGN and star-
forming processes. The discussion above which presents the myriad of
evolutionary scenarios currently postulated serves to illustrate how
much uncertainty remains in this area and the vitally important role
of multi-wavelength datasets in constraining these increasingly
complex evolutionary scenarios.

More complicated modelling of the evolution of radio sources could be
carried out by combining a range of deep fields, e.g. COSMOS
\citep{Smola,Smolb}, SXDF
\citep{Simp2012} and the VLA-VIDEO field in this work and combining surveys with
different flux limits provides much better coverage of the
luminosity-redshift plane \citep[see e.g.][]{Rigby2011}. Future deeper surveys with SKA pathfinder
telescopes, such as MeerKAT will allow us to explore the
the evolution of radio sources over larger areas and to much higher
redshifts ($z\sim$4) overcoming the sample variance limitations
of current small area surveys. 
\label{lastpage}

\section*{Acknowledgments}

The work of Kim McAlpine and Matt Jarvis was supported by the South
African Square Kilometre Array Project and the South African National
Research Foundation. The authors would like to thank the South African
National Research Foundation for funding an intensive research
retreat which made this work possible. Based on observations made with ESO telescopes at the La Silla Paranal Observatory under programme ID:
179.A-2006 and on data products produced
by the Cambridge Astronomy Survey Unit on behalf of the
VIDEO consortium. Based on observations obtained with MegaPrime/MegaCam, a joint project of CFHT and CEA/DAPNIA, at the Canada-France-Hawaii Telescope (CFHT) which is operated by the National Research Council (NRC) of Canada, the Institut National des Science de l'Univers of the Centre National de la Recherche Scientifique (CNRS) of France, and the University of Hawaii. This work is based in part on data products produced at TERAPIX and the Canadian Astronomy Data Centre as part of the Canada-France-Hawaii Telescope Legacy Survey, a collaborative project of NRC and CNRS.

\bibliography{kimsbib.bib}
\bibliographystyle{mn2e}

\end{document}